\documentclass[%
 reprint,
 superscriptaddress,
 amsmath,amssymb,
 prl,
 floatfix,
]{revtex4-1}

\usepackage[colorlinks,
citecolor=red,
urlcolor=blue,
bookmarks=false,
hypertexnames=true]{hyperref} 
\usepackage{color}
\usepackage{graphicx}% Include figure files
\usepackage{dcolumn}% Align table columns on decimal point
\usepackage{bm}% bold math
\begin{document}

\preprint{APS/123-QED}

\title{Single-shot Precise Ranging using Twisted Light}% Force line breaks with \\

\author{Long-Zhu Cen}
\author{Zi-Jing Zhang}
\email{zhangzijing@hit.edu.cn}
\author{Jun-Yan Hu}
\author{Jian-Dong Zhang}
\affiliation{School of Physics, Harbin Institute of Technology, Harbin 150001, China}
\author{Bin Luo}%
\email{luobin@bupt.edu.cn}
\affiliation{State Key Laboratory of Information Photonics and Optical Communications, Beijing University of Posts and Telecommunications, Beijing 100876, China}
\author{Chenglong You}
\author{Omar S. Maga\~na-Loaiza}
\affiliation{Department of Physics and Astronomy, Louisiana State University, Baton Rouge, Louisiana 70803, USA}

\author{Long Wu}
\affiliation{School of informatics, Zhejiang Sci-Tech University, Hangzhou 310018, China}
\author{Yi-Fei Sun}
\author{Yuan Zhao}
\email{zhaoyuan@hit.edu.cn}
\affiliation{School of Physics, Harbin Institute of Technology, Harbin 150001, China}

\date{\today}% It is always \today, today,

\begin{abstract}
Over the past decade, optical orbital angular momentum (OAM) modes were shown to offer advantages in optical information acquisition. 
Here, we introduce a new scheme for optical ranging in which depth is estimated through the angular rotation of petal-like patterns produced by superposition of OAM modes. 
Uncertainty of depth estimation in our strategy depends on how fast the petal-like pattern rotates and how precisely the rotation angle can be estimated.
The impact of these two factors on ranging accuracy are analyzed in presence of noise.
We show that focusing the probe beam provides a quadratic enhancement on ranging accuracy because rotation speed of the beam is inversely proportional to the square of beam radius.
%and study the optimal superposition of OAM modes for a better ranging uncertainty under noises. 
Uncertainty of depth estimation is also proportional to uncertainty of rotation estimation, which can be optimized by picking proper OAM superposition.
Finally, we unveil the possibility of optical ranging for scattering surface with uncertainties of few micrometers under noise.
%This can be achieved through focusing the light beam into a small spot. 
Unlike existing methods which rely on continuous detection for a period of time to achieve such ranging accuracy, our scheme needs only single-shot measurement.
\end{abstract}

\maketitle

%\tableofcontents

%\section{INTRODUCTION}

\textit{Introduction} --- Laser-based light detection and ranging (LIDAR) is a key technology in industrial and scientific metrology, which provides long-range, high-precision, and fast acquisition~\cite{Berkovic_2012}. These schemes have played important roles in a wide variety of applications, including autopilot, industrial process monitoring, or satellite formation flight.
One of the most fundamental and common ranging method used in LIDAR system is ``time-of-flight (TOF)''~\cite{Bosch_2001}. 
Extensive researches have been proposed to improve ranging accuracy of TOF scheme. Among them were approaches exploiting frequency modulation~\cite{Beheim_1985} or amplitude modulation~\cite{Besl_1988}. 
Except for TOF scheme, interferometer configuration can be used with a highly coherent laser to measure sub-wavelength displacements of an object (see, for example,~\cite{Saleh_1991}). Commercialize interferometric distance sensors based on low-coherence sources have been widely used in biological and medical applications~\cite{Drexler_2004}. 
Driven by emerging ultra-fast applications such as real-time measurement of fast non-repetitive events, fast and accurate ranging systems using optical frequency combs~\cite{Udem_2002} have been demonstrated. Exploiting TOF schemes~\cite{Minoshima_2000}, interferometric approaches~\cite{Schuhler_2006}, or combinations thereof~\cite{Coddington_2009}, these new systems have shown characteristic advantages in acquisition rate and accuracy~\cite{Trocha_2018}.

All the schemes mentioned above have taken advantage of light's  temporal coherence, which require inevitable cumulative detection to acquire distance information. On the other hand, single-shot measurement can be achieved by utilizing spatial coherence of light, which is realized based on transverse mode modulation. Over the past decade, one class of the most notable resources associate with transverse mode of light were so-called orbital angular momentum (OAM) modes~\cite{Allen_1992}. 
These modes have recently been used in spinning objects detection~\cite{Barreiro_2006,Lavery_2013,Lavery_2014}, ultra-sensitive angle measurements~\cite{Magana_2014,D_Ambrosio_2013}, imaging~\cite{Torner_2005,Omar_2019}, object identification~\cite{Yang_2017,Uribe_Patarroyo_2013, Zhang_2018}, and quantum pattern recognition~\cite{Qiu_2019}.
%However, ranging scheme based on OAM modes has not been reported yet.

Here we show that OAM superpositions provide another possibility for precise ranging. The principle is similar to that based on frequency modulation or amplitude modulation, while cumulative detection is not required.
The intensity distribution of superposed OAM modes will change during the propagation, acting like a rotating petals. 
By resolving variation of the rotation angle of the petal-like pattern, we demonstrate the ability of single-shot ranging with accuracy up to micron level.

\begin{figure}[tbp]
\centering
\includegraphics[width=\linewidth]{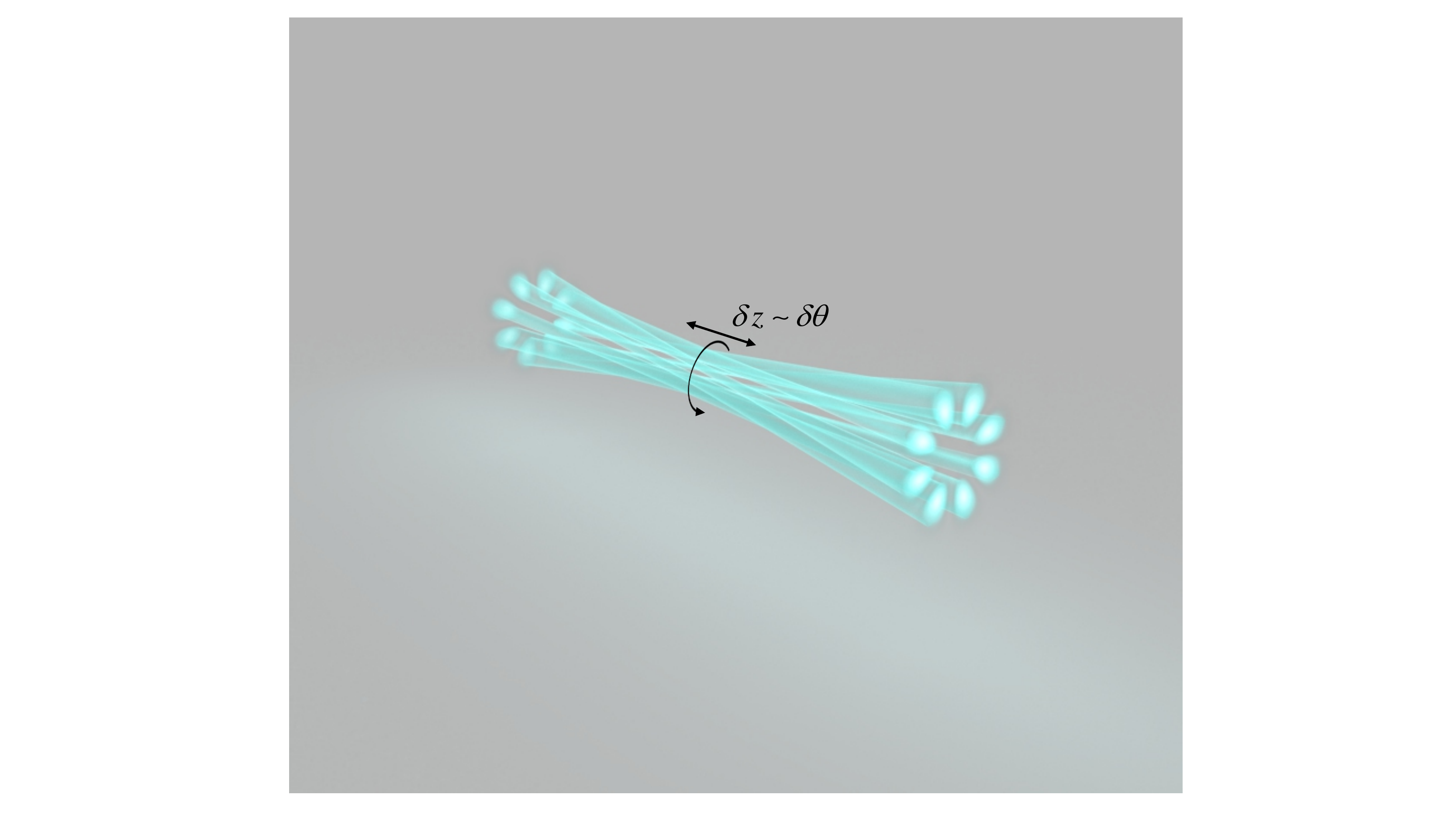}
\caption{Principle of ranging with twisted light produced by superposition of OAM modes. The petal-like intensity distribution rotates as the light propagating forward. Upon knowing the relationship of rotation angle $\theta$ and propagating distance $z$, one may realize ranging of positional displacement $\delta z$ by measuring angular displacement $\delta \theta$.}
\label{fig:scheme}
\end{figure}

%\section{RANGING PRINCIPLE}
\textit{Ranging principle} \---- Laguerre-Gaussian (LG) beams are class of representative laser modes which carrying OAM as highlighted by Allen \emph{et al.}~\cite{Allen_1992}. Here we also adopt LG modes to illuminate the ranging principle based on superposition of OAM modes. Full LG modes are characterized by radial and azimuthal indexes $p$ and $\ell$, while OAM carried by an LG beam is only related to azimuthal index (also called topological charge). Therefore, in our case, we consider a class of simple LG modes whose radial index is zero. Their complex amplitude under cylindrical coordinates $(r, \phi, z)$ can be expressed as~\cite{Allen_1992}:
\begin{align}
\nonumber u_\ell &= \sqrt{\frac {2}{\pi |\ell|!}} \frac{(\sqrt 2\xi)^{|\ell|}}{w(z)}\exp(-\xi^2) \exp ( - i\frac{z}{z_\mathrm{R}}\xi^2 )\\
&\quad \times \exp ( - i\ell\phi )\exp [ i(|\ell|+1) \arctan ( {z}/{z_\mathrm{R}} ) ],
\label{LG}
\end{align}
where $z_\mathrm{R}$ is Rayleigh length and $\xi = r/w(z)$ with $w(z) = [ 2(z^2 + z^2_\mathrm{R})/(kz_\mathrm{R})]^{1/2}$ being the beam radius at $z$ ($k$ is the wave number). Superposition of LG beams is extensively discussed in ~\cite{Kulkarni_2017}. As for two beams with arbitrary topological charges $\ell_1,\ell_2$, complex amplitude of the superposition can be denoted as $u_{\ell_1,\ell_2}=\rho_{\ell_1}e^{iS_{\ell_1}}+\rho_{\ell_2}e^{iS_{\ell_2}}$, where $\rho_{\ell}$ and $S_{\ell}$ represent the amplitude and phase in Eq.~(\ref{LG}), respectively. Then, the intensity distribution can be obtained as the square modulus:
\begin{equation}
I_{\ell_1,\ell_2}(r, \phi, z)=\rho_{\ell_1}^2 + \rho_{\ell_2}^2 + 2\rho_{\ell_1}\rho_{\ell_2}\cos(S_{\ell_1} - S_{\ell_2}).
\end{equation}
The intensity distribution shows a petal-like pattern which is rotational recognizable when $\ell_1 \ne \ell_2$. Additionally, if one compares the intensity distribution at $z=0$ with that of any other position, it is not difficult to find out that their expressions satisfy:
\begin{equation}
I_{\ell_1,\ell_2}(r, \phi, z)= \alpha^2I_{\ell_1,\ell_2}(\alpha r, \phi+\theta, 0).
\end{equation}
The parameter $\alpha = w(0)/w(z)$ is radial scaling ratio and
\begin{equation}
\theta = \Delta |\ell|\arctan(z/z_\mathrm R)/\Delta \ell
\label{angle_vs_distance}
\end{equation}
is rotation angle of the intensity pattern at $z$ with respect to that of $z=0$, where $\Delta |\ell| = |\ell_2| - |\ell_1|$ and $\Delta \ell = \ell_1 - \ell_2$. Equation~(\ref{angle_vs_distance}) shows that $|\theta|$ increases with the increase of $z$ as long as $\Delta |\ell| \ne 0$, which means superposition of such two modes produce a rotating intensity pattern. This particular feature makes OAM superpositions useful tools for estimating positional displacement $\delta z$ by measuring angular displacement $\delta \theta$, as shown in Fig.~\ref{fig:scheme}.

A convenient method for rotation estimation is needed to help realizing depth measurement by OAM superpositions. As for this problem, previous researches on rotating field adopt rather direct ways. Mostly rely on measuring the angular displacement of symmetry axis with the help of centroid algorithm~\cite{Schulze_2015, Webster_2017}. These methods need adjustment for different patterns and are not suitable for a universal estimation process. Here we show another method based on circular harmonic expansion which was firstly used in pattern recognition~\cite{Hsu_1982_rotation}. 

A reference which determines the original orientation can be obtained by decomposing the intensity pattern at $z = 0$ into circular harmonic components
\begin{equation}
I_{\ell_1,\ell_2}(r, \phi, 0)=\sum\limits_{M =  - \infty }^{ + \infty } f_M(r)\exp(iM\phi),
\end{equation}
where the expansion coefficient 
\begin{equation}
f_M(r)=\frac{1}{2\pi}\int_0^{2\pi} I_{\ell_1,\ell_2}(r, \phi, 0)\exp(-iM\phi)d\phi.
\end{equation}
Given the expression of $I_{\ell_1,\ell_2}|_{z=0}$, $f_M(r)$ turns out to be:
\begin{equation}
f_M(r)=\left\{ {
\begin{aligned}
\rho_{\ell_1}\rho_{\ell_2}|_{z=0},\;\quad\quad &M = \pm(\ell_1-\ell_2)\\
\rho_{\ell_1}^2|_{z=0} + \rho_{\ell_2}^2|_{z=0},\quad &M = 0\\
0,\quad\quad\quad\quad &\mathrm{Others}
\end{aligned}
} \right.
\label{CHEcoefficient}
\end{equation}
In order to measure the rotation angle $\theta$ of $I_{\ell_1,\ell_2}(r, \phi, z)$ with respect to $I_{\ell_1,\ell_2}(r, \phi, 0)$, one can choose one of the circular harmonic components with nonzero $M$,
\begin{equation}
F_M(r,\phi)=f_M(r)\exp(iM\phi),
\label{reference}
\end{equation}
as a reference, then, the cross-correlation function of $I_{\ell_1,\ell_2}(r, \phi, z)$ and $F_M(r,\phi)$ in Cartesian coordinates is
\begin{equation}
R_M(x,y)=\iint_{-\infty}^{\infty}I_{\ell_1,\ell_2}(\xi, \eta, z)F_M^*(\xi - x, \eta - y)d\xi d\eta.
\end{equation}
The center correlation, $R_M(0,0)$, is the value at the origin of the 2-D correlation function $R_M(x,y)$. It possesses the maximal modulus and can be calculated also in polar coordinates $(r, \phi)$ whose origin coincides with that of the Cartesian coordinates. Let $C_M$ denote $R_M(0,0)$, we obtain
\begin{align}
\nonumber C_M&=\int_{0}^{\infty}rdr\int_0^{2\pi}I_{\ell_1,\ell_2}(\alpha r, \phi+\theta, 0)F_M^*(r, \phi)d\phi\\
&=A\exp(iM\theta),
\label{CenterCorrelation_vs_angle}
\end{align}
with $A=2\pi\int_{0}^{\infty}f_M(\alpha r)f_M^*(r)rdr$. Notice that $\rho_\ell$ is real and so is $f_M(r)$ and $A$, the rotation angle $\theta$ can then be obtained from the phase of $C_M$:
\begin{equation}
\theta = \frac{\mathrm{arg} [C_M]}{M} = \frac{\Theta}{M} .
\label{angle_vs_phase}
\end{equation}

Although radial scaling ratio between reference and detected pattern $\alpha$ does not affect the phase of center correlation $C_M$ according to Eq.~(\ref{CenterCorrelation_vs_angle}), it takes contribute into the amplitude $A$. The amplitude decreases with the increase of $\alpha$, which makes the measurement of rotating angle impressionable under inevitable disturbance, as we shall show what noise does in realistic measurement. However, the influence of $\alpha$ is not difficult to eliminate since the scale of reference can be chosen arbitrarily to meet that of the detected pattern. One can easily find the proper size of reference by some scale estimation algorithms such as fitting the detected pattern to a standard one. Because of this, we will take no account the zoom of pattern scale and set $\alpha = 1$ to obtain that
\begin{equation}
A \propto \frac{\Gamma(|\ell_1| + |\ell_2| + 1)}{\Gamma(|\ell_1| + 1)\Gamma(|\ell_2| + 1)2^{|\ell_1| + |\ell_2|}}.
\label{amplitude}
\end{equation}

With rotation angle obtained, displacement of the target should be sensed by monitoring the variation of rotation angle, and the exact value can be calculated by Eq.~(\ref{angle_vs_distance}). However, ranging accuracy of such scheme still remain to be discussed. We will be concerned with this question in the rest of this Letter.

%\section{Ranging accuracy in realistic measurement}
\textit{Ranging accuracy in realistic measurement} --- In realistic scenes, experimental uncertainties exist in the implementation of the scheme and eventually result in ranging error. In our case, finite resolution and background noise of detecting device appear to be the main adverse factors. Hence, we discuss ranging accuracy with these two factors taken into consideration, and show the optimal superposition of OAM modes under certain noise level.

Assume that the detecting device has a identical resolution $\Delta L$ in both $x$ and $y$ directions. The intensity distribution functions of detected patterns are  discretized and can be written as $I_{\ell_1,\ell_2}(r_{nm}, \phi_{nm}, z)$, where $r_{nm} = \Delta L\sqrt{n^2 + m^2}$, $\tan\phi_{nm} = m/n$ with $n,m = 0, \pm1, \pm2, \pm3, \cdots$ are indexs in $x$ and $y$ directions respectively. Then, the reference should also be discretized as $F_M(r_{nm}, \phi_{nm})$ to perform discrete cross-correlation with $I_{\ell_1,\ell_2}(r_{nm}, \phi_{nm}, z)$, which makes Eq.~(\ref{CenterCorrelation_vs_angle}) become:
\begin{equation}
C_M'=(\Delta L)^2\sum\limits_{n,m} I_{\ell_1,\ell_2}(r_{nm}, \phi_{nm}, z)F_M^*(r_{nm}, \phi_{nm})
\end{equation}
Let $A'$ be the module of $C_M'$. According to Riemann integral, $C_M'$ and $A'$ will converge to $C_M$ and $A$ respectively as $\Delta L$ turns to zero. In fact, the differences between values with and without a prime in our case become as small as $10^{-10}$ when $\Delta L = w(z)/300$ according to numerical estimation, which is easy to meet by a common imaging system~\footnote{As for light spot with radius of micron scale and commercial CCD or CMOS whose pixel size is also several micrometers, one may need a microscope system with over 300$\times$ magnification to make the errors result from finite resolution smaller than $10^{-10}$}. Therefore, errors result only from finite resolution are negligible. It is the random noise in each pixel originating from background light or dark current that affects the measured result mostly.

To understand how noise affects the ranging process, we substitute Eq.(\ref{angle_vs_phase}) into Eq.~(\ref{angle_vs_distance}) and differentiate both sides of the resulted equation to obtain that:
\begin{equation}
\Delta z = \frac{\pi w^2(z)}{\Delta |\ell|\lambda }\Delta \Theta,
\label{z_uncertainty}
\end{equation}
which indicates that ranging accuracy, quantified by uncertainty of estimation $\Delta z$, is proportional to the square of beam radius $w(z)$ at where the target is placed. Hence, one can enhance ranging accuracy very effectively by focusing the beam.
%~\footnote{The focus process can be accomplished by one microscope system which also magnifies the pattern on target for imaging, with the same magnification of focusing. Hence, the focus process will not affect the size of the pattern on imaging plane theoretically and consequently will not affect the angle estimation process} 
On the other hand, ranging uncertainty will increase rapidly while the light propagating forward and beam radius increasing. As a result, the optimal working point is around the focus where the beam radius achieves the minimum. Also, to figure out where the exact level of ranging accuracy stands, uncertainty of the estimated angle $\Delta \Theta$ should be specified. In the absence of noise, the center correlation is a determined value and ranging will be preformed with no error. The situation when noise exist will be discussed in the following.

We assume that the intensity distribution at position $z$ is detected along with additive Gaussian white noise in each pixel, which results in a noisy pattern $I_{\ell_1,\ell_2}(r_{nm}, \phi_{nm}, z) + N_{nm}$. $\{N_{nm}\}$ are independent random variables that have normal distribution with same mean $\mu$ and variance $\sigma ^2$. Hence, the center correlation $C_M'$ between such random pattern and reference should also be random variable. We show in Appendix that $C_M'$ have normal distribution on the complex plane. Both of it's real and image part have a variance of $A'\sigma ^2/2$ and are centered at the original points when noise is absent. Let $\Theta_0$ be the original phase of $C_M'$, the probability distribution function of phase fluctuation $\tilde\Theta = \Theta - \Theta_0$ is
\begin{align}
\nonumber f_{\tilde\Theta}(\tilde\theta)&=\gamma \sqrt {\frac{A'}{\pi\sigma ^2}} \exp \left({ - \frac{A'\sin^2\tilde \theta}{\sigma ^2}}\right )\cos \tilde \theta\\
&\quad + \frac{1}{2\pi}\exp \left ({- \frac{A'}{\sigma ^2}} \right ),
\label{probability}
\end{align}
where $\gamma  = [\rm{erf}(\sqrt {A'} \cos \tilde \theta /\sigma ) + 1]/2$. Expending $f_{\tilde\Theta}(\tilde\theta)$ at $\tilde\theta = 0$ and ignoring terms higher than the second order, the probability distribution function reduce to Gaussian-type with variance of $\sigma ^2/2A'$.  
It is notable that although Eq.~(\ref{probability}) is obtained when noise has normal distribution, similar results can also be derived for white noise with other distributions according to central-limit theorem. 
Difference among results under different kind of noises is the variance for $\tilde\Theta$.

\begin{figure}[tbp]
\centering
\includegraphics[width=\linewidth]{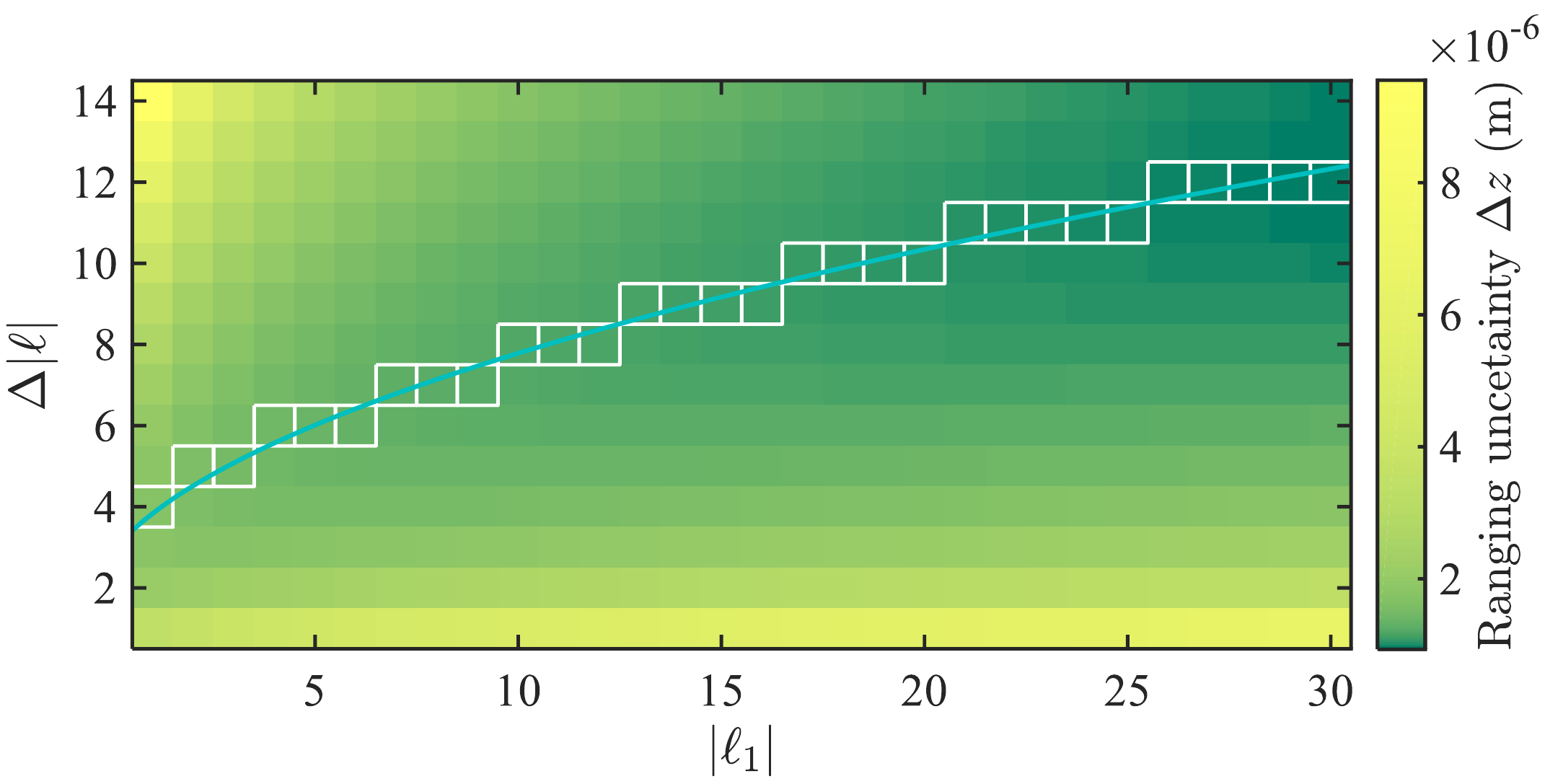}
\caption{Ranging uncertainty under different superposition of OAM modes with topological charges of $\ell_1$ and $\ell_2 = \ell_1 + \Delta |\ell|$. Uncertainty of the estimation is calculated at the focus of an aplanatic objective with a working distance of 10 mm, by which the 532 nm green light with a waist radius of 0.5mm is focused into a spot of 10 $\mu$m. The variance of noise $\sigma^2$ stays constant while equaling to the average of the light's intensity. The solid curve indicates the theoretical optimal $\Delta |\ell|$ with different $\ell_1$ and the actual optimal points are marked by white squares. For noises with different variances, the optimal superposition keeps the same, while the exact ranging uncertainty different. The optimal ranging uncertainty under the above conditions reaches 1 $\mu$m.}
\label{fig:ranging_accuracy}
\end{figure}

Since $\Theta$ differs with $\tilde \Theta$ only by a constant value, uncertainty of the phase $\Delta \Theta$ equals to that of the phase fluctuation $\sigma/\sqrt{2A'}$. Then, according to Eq.~(\ref{z_uncertainty}), one may find out that the higher $\sqrt{A'}$ is, the smaller the uncertainty of distance estimation is. Since $A'$ decreases with respect to the rise of $\Delta |\ell|$, the optimal ranging accuracy under noise with certain variance appears when $\Delta |\ell|\sqrt{A'}$ reaches the maximum. We replace $|\ell_2|$ with $|\ell_1| + \Delta |\ell|$ in Eq.~(\ref{amplitude}) and solve $\partial(\Delta |\ell|\sqrt{A})/\partial(\Delta |\ell|)=0$ to find out the optimal $\Delta |\ell|$. Figure~\ref{fig:ranging_accuracy} shows optimal $\Delta |\ell|$ for different lower-order topological charges $|\ell_1|$. The optimal ranging uncertainty in our calculation reaches 1 $\mu$m. This is achieved by focusing a light beam at 532 nm into a spot with a radius of 10 $\mu$m when variance of the noise is equal to average intensity of the spot. 
Compared with previous technology~\cite{Trocha_2018} which also achieves such ranging accuracy, our scheme realizes single-shot measurement instead of continuous detection for a period of time. Combined with a ultra high speed camera~\cite{Liang_2018}, the acquisition rate can be improved by a factor of $10^4$.The fundamental limit lies in our scheme is the resolution of optical imaging system. Rotation angle of the pattern will be impossible to estimate if the imaging system can not distinguish each petal in the spot. This may happen when the beam is focused too small or there are too many petals in one spot because $\Delta |\ell|$ is too large.

\textit{Conclusions} \---- Our paper elucidates potential ability of OAM superpositions for precise ranging, which is internally owing to rotating nature of the intensity distribution during light's propagation. 
%Distance can be estimated through the angular rotation of petal-like patterns.
Our scheme provide an effective method to enhance ranging accuracy by focusing the beam into a small spot. 
We have shown that uncertainty of depth estimation goes as the square of the beam radius at where the target is placed. 
This is mainly because rotation speed of the beam grows quadratically as the beam radius goes down, which makes the intensity pattern rotates more sensitively with the variation of distance.
%Just like the relation between angular velocity and rotational inertia in a system whose angular momenta conserve. 
Our scheme also does not relies on continuous detection to achieve high ranging accuracy.
In this regard, ranging for scattering surface with uncertainties of few micrometers can be realized by single-shot manner.
Choice among all possible OAM superpositions is crucial because specific distributions of the petal-like patterns affect the accuracy of rotation estimation. 
In general, the greater the difference between absolute values of two topological charges $\Delta |\ell|$ is, the faster the patterns rotate, which is good for improving the ranging accuracy. 
However, the increase of $\Delta |\ell|$ also fades the petal-like patterns and increase the errors in rotation measurement under noise. 
Hence there exist optimal choices for  $\Delta |\ell|$ to trade-off the advantages and disadvantages. 
We have derived analytical equation for finding the optimal $\Delta |\ell|$ if the noise involved in the measurement process are Gaussian white noise. 
As the lower-order topological increase, the optimal $\Delta |\ell|$ becomes greater, and the final ranging uncertainty goes down.

%OAM superpositions can be used for precise single-shot ranging. 
The scheme presented in this article actually acquires the average depth of the target surface by estimation the average rotation angle of the whole intensity pattern. And this is accomplished by a camera and data processing on computer, which calls for high demand on speed of the camera and processor in ultrafast application. In fact, there is alternative method to improve the acquisition rate which adopts optical means~\cite{Hsu_1982} for rotation estimation.
Except for average depth, the pattern distortion resulted from uneven target actually reveals the micro surface topography of the illuminated area. Comparing the distorted pattern with a standard one, the depth structure of target surface can be reconstructed by solving the inverse problem with the help of existing methods such as SPIRAL-TAP solver~\cite{Harmany_2012}. We believe that this might pave the way
towards a new method for 3D imaging in a single-shot or a few-shots manner.

%Some useful techniques may be helpful in realistic implementation. For example, the acquisition rate can be greatly improved by adopting optical means~\cite{Hsu_1982} to estimate the rotation.  can be used to solve the inverse problems commonly exist in processes of depth reconstruction.
%Since the light keeps rotating after the reflection, without additional setup, the rotational speed of the pattern detected by the imaging device will not be what it is at the waist, and so is the sensitivity. 
%Can angle measurement be achieved by faster optical means?  How far dose the quality of OAM modes affects the ranging accuracy? These questions are worthy of research while superposed OAM modes proving itself useful in practical ranging application.

\textit{Acknowledgment} \---- We thank B.-W. G. for helpful discussions.

L.-Z. C. and Z.-J. Z. contributed equally to this work.

\begin{widetext}

\appendix*

\section{APPENDIX}

Here we prove the independence of real and image parts of center correlation between reference and detected pattern under Gaussian white noise. Based on this, we show how to calculate the probability distribution of phase fluctuation.
\begin{figure}[bp]
\centering
\includegraphics[width=0.5\linewidth]{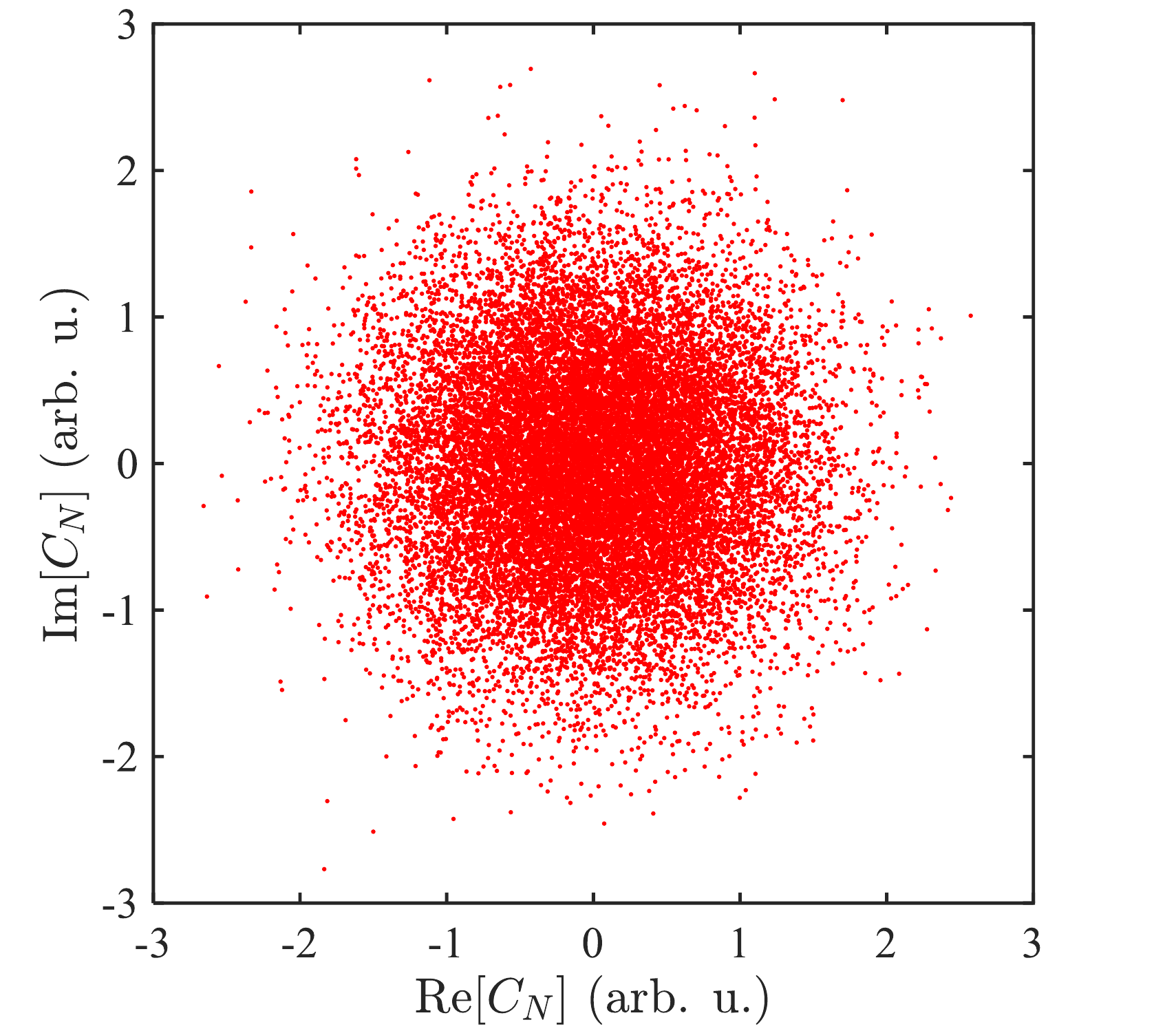}
\caption{Distribution of center correlation between noise and reference $C_N$ on complex plane. The pair of random variables, $\mathrm{Re}[C_N]$ and $\mathrm{Im}[C_N]$, has a bivariate normal distribution with zero covariance, which implies the independence of them.}
\label{fig:CN distribution}
\end{figure}

For noisy detected pattern $I_{\ell_1,\ell_2}(r_{nm}, \phi_{nm}, z) + N_{nm}$, the center correlation with reference $F_M (r_{nm}, \phi_{nm})$ becomes $C_M' + C_N$ with $C_M'=A'\exp(i\Theta_0)$ being the constant signal term and $C_N$ being the fluctuant noise term:
\begin{equation}
C_N=(\Delta L)^2\sum\limits_{n,m} N_{nm}F_M^*(r_{nm}, \phi_{nm},).
\end{equation}
Using the expressions for the reference (main text Eq.~(7-8)) we find
\begin{subequations}
\begin{eqnarray}
\mathrm{Re}[C_N]&=&(\Delta L)^2\sum\limits_{n,m} N_{nm}f_M(r_{nm})\cos(M\phi_{nm}),
\\
\mathrm{Im}[C_N]&=&(\Delta L)^2\sum\limits_{n,m} N_{nm}f_M(r_{nm})\sin(M\phi_{nm}),
\end{eqnarray}
\end{subequations}
Since $N_{nm}$ are independent random variables that have normal distribution with the same mean $\mu$ and variance $\sigma^2$, it can be concluded that real and image parts of $C_N$ (and any linear combination of them) are also normal distributed random variables because they are linear transformation of $\{N_{nm}\}$, which leads to bivariate normal distribution of $C_N$ on complex plane as shown in Fig.~\ref{fig:CN distribution}. 

For this pair of random variables, they are independent if they are uncorrelated~\cite{Bartoszy_ski_2007}, which is true in our case because the  covariance of $\mathrm{Re}[C_N]$ and $\mathrm{Im}[C_N]$ goes to zero as $\Delta L$ turns to zero. To prove this, we adopt $X$ and $Y$ to represent $\mathrm{Re}[C_N]$ and $\mathrm{Im}[C_N]$ for concision. 

By definition, the covariance of $X$ and $Y$ is $\mathrm{cov}(X, Y)=\mathrm{E}(XY)-\mathrm{E}(X)\mathrm{E}(Y)$. As for $\mathrm{E}(X)$, we find
\begin{equation}
\mathrm{E}(X) = (\Delta L)^2\sum\limits_{n,m} \mathrm{E}(N_{nm})f_M(r_{nm})\cos(M\phi_{nm})
=\mu (\Delta L)^2\sum\limits_{n,m} f_M(r_{nm})\cos(M\phi_{nm}) = 0,
\label{EX}
\end{equation}
in which we use integral to replace summation while $\Delta L$ turns to zero. $\mathrm{E}(Y)$ can be similarly proved to be zero. 

As for $\mathrm{E}(XY)$, we find
\begin{align}
\nonumber \mathrm{E}(XY) &= (\Delta L)^4\sum\limits_{n,m,j,k} \mathrm{E}(N_{nm}N_{jk})f_M(r_{nm})f_M(r_{jk})\cos(M\phi_{nm})\sin(M\phi_{jk})
\\
\nonumber &=(\Delta L)^4\sum\limits_{n,m,j,k }^{n \ne j\& m \ne k} \mathrm{E}(N_{nm}N_{jk})f_M(r_{nm})f_M(r_{jk})\cos(M\phi_{nm})\sin(M\phi_{jk})
\\
& \quad +(\Delta L)^4\sum\limits_{n,m} \mathrm{E}(N_{nm}^2) f_M^2(r_{nm})\cos(M\phi_{nm})\sin(M\phi_{nm}).
\label{EXY1}
\end{align}
The expectation of the product of two independent random variables equals the product of the expectation of each variable. Hence, $\mathrm{E}(N_{nm}N_{jk}) = \mu^2$ while $n \ne j \& m \ne k$. Moreover, the square of a normal distributed random variable is a chi-square distribution with one degree of freedom, whose expectation equals $1 + \mu^2$. As a result, Eq.~(\ref{EXY1}) turns to:
\begin{align}
\nonumber \mathrm{E}(XY) &=\mu^2(\Delta L)^4\sum\limits_{n,m,j,k }^{n \ne j\& m \ne k} f_M(r_{nm})f_M(r_{jk})\cos(M\phi_{nm})\sin(M\phi_{jk})\\
\nonumber & \quad +(1 + \mu^2)(\Delta L)^4\sum\limits_{n,m} f_M^2(r_{nm})\cos(M\phi_{nm})\sin(M\phi_{nm})
\\
\nonumber &=\mu^2(\Delta L)^4\sum\limits_{n,m,j,k } f_M(r_{nm})f_M(r_{jk})\cos(M\phi_{nm})\sin(M\phi_{jk})\\
\nonumber & \quad +(\Delta L)^4\sum\limits_{n,m} f_M^2(r_{nm})\cos(M\phi_{nm})\sin(M\phi_{nm})\\
&=0.
\label{EXY2}
\end{align}
From Eq.~(\ref{EX}) and~(\ref{EXY2}) we obtain $\mathrm{cov}(X, Y)=0$ while $\Delta L$ turns to zero. Hence the real and image parts of center correlation between reference and detected pattern under Gaussian white noise can be regarded independent when the difference between integral and summation is negligible, which is true in our case as mentioned in the main text.

The probability distribution function of $C_N$ on complex plane $f_{XY}$ equals the product of probability distribution functions of $X$ and $Y$, $f_Xf_Y$, when $X$ and $Y$ are independent. As mentioned above, $X$ and $Y$ are normally distributed with zero expectation. To obtain the probability distribution function of $X$ and $Y$, one has to know the variance of them. As the linear transformation of independent normal random variables $\{N_{nm}\}$, the variance of $X$ and $Y$ can be calculated as:
\begin{equation}
\mathrm{D}(X) =(\Delta L)^2\sum\limits_{n,m} \mathrm{D}(N_{nm}) f_M^2(r_{nm})\cos^2(M\phi_{nm})
=\sigma^2(\Delta L)^2\sum\limits_{n,m}f_M^2(r_{nm})\cos^2(M\phi_{nm})=A'\sigma^2/2,
\label{DX}
\end{equation}
where $A'=2\pi(\Delta L)^2\sum_{n,m} f_M^2(r_{nm})$ is the same as in the main text while $\alpha=1$. The same result can be obtained for the case of $Y$. 

Let $X'$ and $Y'$ represent the real and image parts of the center correlation between $I_{l_1,l_2}(n\Delta l, m\Delta l, z) + N_{nm}$ and reference. Then we obtain $X'= X + A'\cos\Theta_0$ and $Y'= Y + A'\sin\Theta_0$ and the probability distribution function of the center correlation on complex plane can be written as:
\begin{equation}
f_{X'Y'}(x,y)=\frac{1}{2\pi}\exp\left [ \frac{2(x-A'\cos\Theta_0)^2}{A'\sigma^2} + \frac{2(y-A'\sin\Theta_0)^2}{A'\sigma^2} \right].
\end{equation}

Since we concern fluctuation of the phase of center correlation, we obtain the probability distribution function of module $R$ and phase $\Theta$ by transformation~\cite{Devore_2012}:
\begin{equation}
f_{R\Theta}(r,\theta)=rf_{X'Y'}(r\cos\theta, r\sin\theta).
\end{equation}
Then the marginal distribution of $\Theta$ can be obtained by integrating $f_{R\Theta}(r,\theta)$ with respect to $r$, which results in:
\begin{equation}
f_{\tilde\Theta}(\tilde\theta)=
\frac{1}{2\pi}\exp \left ({- \frac{A'}{\sigma ^2}} \right )
+\gamma \sqrt {\frac{A'}{\pi\sigma ^2}} \exp \left({ - \frac{A'\sin^2\tilde \theta}{\sigma ^2}}\right )\cos \tilde \theta,
\end{equation}
where $\gamma  = [\mathrm{erf}(\sqrt {A'} \cos \tilde \theta /\sigma ) + 1]/2$ and $\tilde\Theta = \Theta - \Theta_0$ represents the phase fluctuation.
\end{widetext}


\bibliography{apssamp}% Produces the bibliography via BibTeX.

\begin{thebibliography}{33}%
\makeatletter
\providecommand \@ifxundefined [1]{%
 \@ifx{#1\undefined}
}%
\providecommand \@ifnum [1]{%
 \ifnum #1\expandafter \@firstoftwo
 \else \expandafter \@secondoftwo
 \fi
}%
\providecommand \@ifx [1]{%
 \ifx #1\expandafter \@firstoftwo
 \else \expandafter \@secondoftwo
 \fi
}%
\providecommand \natexlab [1]{#1}%
\providecommand \enquote  [1]{``#1''}%
\providecommand \bibnamefont  [1]{#1}%
\providecommand \bibfnamefont [1]{#1}%
\providecommand \citenamefont [1]{#1}%
\providecommand \href@noop [0]{\@secondoftwo}%
\providecommand \href [0]{\begingroup \@sanitize@url \@href}%
\providecommand \@href[1]{\@@startlink{#1}\@@href}%
\providecommand \@@href[1]{\endgroup#1\@@endlink}%
\providecommand \@sanitize@url [0]{\catcode `\\12\catcode `\$12\catcode
  `\&12\catcode `\#12\catcode `\^12\catcode `\_12\catcode `\%12\relax}%
\providecommand \@@startlink[1]{}%
\providecommand \@@endlink[0]{}%
\providecommand \url  [0]{\begingroup\@sanitize@url \@url }%
\providecommand \@url [1]{\endgroup\@href {#1}{\urlprefix }}%
\providecommand \urlprefix  [0]{URL }%
\providecommand \Eprint [0]{\href }%
\providecommand \doibase [0]{http://dx.doi.org/}%
\providecommand \selectlanguage [0]{\@gobble}%
\providecommand \bibinfo  [0]{\@secondoftwo}%
\providecommand \bibfield  [0]{\@secondoftwo}%
\providecommand \translation [1]{[#1]}%
\providecommand \BibitemOpen [0]{}%
\providecommand \bibitemStop [0]{}%
\providecommand \bibitemNoStop [0]{.\EOS\space}%
\providecommand \EOS [0]{\spacefactor3000\relax}%
\providecommand \BibitemShut  [1]{\csname bibitem#1\endcsname}%
\let\auto@bib@innerbib\@empty
%</preamble>
\bibitem [{\citenamefont {Berkovic}\ and\ \citenamefont
  {Shafir}(2012)}]{Berkovic_2012}%
  \BibitemOpen
  \bibfield  {author} {\bibinfo {author} {\bibfnamefont {G.}~\bibnamefont
  {Berkovic}}\ and\ \bibinfo {author} {\bibfnamefont {E.}~\bibnamefont
  {Shafir}},\ }\href {\doibase 10.1364/Aop.4.000441} {\bibfield  {journal}
  {\bibinfo  {journal} {Advances in Optics and Photonics}\ }\textbf {\bibinfo
  {volume} {4}},\ \bibinfo {pages} {441} (\bibinfo {year} {2012})}\BibitemShut
  {NoStop}%
\bibitem [{\citenamefont {Bosch}(2001)}]{Bosch_2001}%
  \BibitemOpen
  \bibfield  {author} {\bibinfo {author} {\bibfnamefont {T.}~\bibnamefont
  {Bosch}},\ }\href {\doibase 10.1117/1.1330700} {\bibfield  {journal}
  {\bibinfo  {journal} {Optical Engineering}\ }\textbf {\bibinfo {volume}
  {40}},\ \bibinfo {pages} {10} (\bibinfo {year} {2001})}\BibitemShut {NoStop}%
\bibitem [{\citenamefont {Beheim}\ and\ \citenamefont
  {Fritsch}(1985)}]{Beheim_1985}%
  \BibitemOpen
  \bibfield  {author} {\bibinfo {author} {\bibfnamefont {G.}~\bibnamefont
  {Beheim}}\ and\ \bibinfo {author} {\bibfnamefont {K.}~\bibnamefont
  {Fritsch}},\ }\href {\doibase 10.1049/el:19850064} {\bibfield  {journal}
  {\bibinfo  {journal} {Electronics Letters}\ }\textbf {\bibinfo {volume}
  {21}},\ \bibinfo {pages} {93} (\bibinfo {year} {1985})}\BibitemShut {NoStop}%
\bibitem [{\citenamefont {Besl}(1988)}]{Besl_1988}%
  \BibitemOpen
  \bibfield  {author} {\bibinfo {author} {\bibfnamefont {P.~J.}\ \bibnamefont
  {Besl}},\ }\href {\doibase 10.1007/bf01212277} {\bibfield  {journal}
  {\bibinfo  {journal} {Machine Vision and Applications}\ }\textbf {\bibinfo
  {volume} {1}},\ \bibinfo {pages} {127} (\bibinfo {year} {1988})}\BibitemShut
  {NoStop}%
\bibitem [{\citenamefont {Saleh}\ and\ \citenamefont
  {Teich}(1991)}]{Saleh_1991}%
  \BibitemOpen
  \bibfield  {author} {\bibinfo {author} {\bibfnamefont {B.~E.~A.}\
  \bibnamefont {Saleh}}\ and\ \bibinfo {author} {\bibfnamefont {M.~C.}\
  \bibnamefont {Teich}},\ }\href {\doibase 10.1002/0471213748} {\emph {\bibinfo
  {title} {Fundamentals of Photonics}}}\ (\bibinfo  {publisher} {John Wiley
  {\&} Sons, Inc.},\ \bibinfo {year} {1991})\BibitemShut {NoStop}%
\bibitem [{\citenamefont {Drexler}(2004)}]{Drexler_2004}%
  \BibitemOpen
  \bibfield  {author} {\bibinfo {author} {\bibfnamefont {W.}~\bibnamefont
  {Drexler}},\ }\href {https://doi.org/10.1117%2F1.1629679} {\bibfield
  {journal} {\bibinfo  {journal} {Journal of Biomedical Optics}\ }\textbf
  {\bibinfo {volume} {9}},\ \bibinfo {pages} {47} (\bibinfo {year}
  {2004})}\BibitemShut {NoStop}%
\bibitem [{\citenamefont {Udem}\ \emph {et~al.}(2002)\citenamefont {Udem},
  \citenamefont {Holzwarth},\ and\ \citenamefont {Hänsch}}]{Udem_2002}%
  \BibitemOpen
  \bibfield  {author} {\bibinfo {author} {\bibfnamefont {T.}~\bibnamefont
  {Udem}}, \bibinfo {author} {\bibfnamefont {R.}~\bibnamefont {Holzwarth}}, \
  and\ \bibinfo {author} {\bibfnamefont {T.~W.}\ \bibnamefont {Hänsch}},\
  }\href {\doibase 10.1038/416233a} {\bibfield  {journal} {\bibinfo  {journal}
  {Nature}\ }\textbf {\bibinfo {volume} {416}},\ \bibinfo {pages} {233}
  (\bibinfo {year} {2002})}\BibitemShut {NoStop}%
\bibitem [{\citenamefont {Minoshima}\ and\ \citenamefont
  {Matsumoto}(2000)}]{Minoshima_2000}%
  \BibitemOpen
  \bibfield  {author} {\bibinfo {author} {\bibfnamefont {K.}~\bibnamefont
  {Minoshima}}\ and\ \bibinfo {author} {\bibfnamefont {H.}~\bibnamefont
  {Matsumoto}},\ }\href {\doibase 10.1364/ao.39.005512} {\bibfield  {journal}
  {\bibinfo  {journal} {Applied Optics}\ }\textbf {\bibinfo {volume} {39}},\
  \bibinfo {pages} {5512} (\bibinfo {year} {2000})}\BibitemShut {NoStop}%
\bibitem [{\citenamefont {Schuhler}\ \emph {et~al.}(2006)\citenamefont
  {Schuhler}, \citenamefont {Salvad{\'{e}}}, \citenamefont
  {L{\'{e}}v{\^{e}}que}, \citenamefont {Dändliker},\ and\ \citenamefont
  {Holzwarth}}]{Schuhler_2006}%
  \BibitemOpen
  \bibfield  {author} {\bibinfo {author} {\bibfnamefont {N.}~\bibnamefont
  {Schuhler}}, \bibinfo {author} {\bibfnamefont {Y.}~\bibnamefont
  {Salvad{\'{e}}}}, \bibinfo {author} {\bibfnamefont {S.}~\bibnamefont
  {L{\'{e}}v{\^{e}}que}}, \bibinfo {author} {\bibfnamefont {R.}~\bibnamefont
  {Dändliker}}, \ and\ \bibinfo {author} {\bibfnamefont {R.}~\bibnamefont
  {Holzwarth}},\ }\href {\doibase 10.1364/ol.31.003101} {\bibfield  {journal}
  {\bibinfo  {journal} {Optics Letters}\ }\textbf {\bibinfo {volume} {31}},\
  \bibinfo {pages} {3101} (\bibinfo {year} {2006})}\BibitemShut {NoStop}%
\bibitem [{\citenamefont {Coddington}\ \emph {et~al.}(2009)\citenamefont
  {Coddington}, \citenamefont {Swann}, \citenamefont {Nenadovic},\ and\
  \citenamefont {Newbury}}]{Coddington_2009}%
  \BibitemOpen
  \bibfield  {author} {\bibinfo {author} {\bibfnamefont {I.}~\bibnamefont
  {Coddington}}, \bibinfo {author} {\bibfnamefont {W.~C.}\ \bibnamefont
  {Swann}}, \bibinfo {author} {\bibfnamefont {L.}~\bibnamefont {Nenadovic}}, \
  and\ \bibinfo {author} {\bibfnamefont {N.~R.}\ \bibnamefont {Newbury}},\
  }\href {\doibase 10.1038/nphoton.2009.94} {\bibfield  {journal} {\bibinfo
  {journal} {Nature Photonics}\ }\textbf {\bibinfo {volume} {3}},\ \bibinfo
  {pages} {351} (\bibinfo {year} {2009})}\BibitemShut {NoStop}%
\bibitem [{\citenamefont {Trocha}\ \emph {et~al.}(2018)\citenamefont {Trocha},
  \citenamefont {Karpov}, \citenamefont {Ganin}, \citenamefont {Pfeiffer},
  \citenamefont {Kordts}, \citenamefont {Wolf}, \citenamefont {Krockenberger},
  \citenamefont {Marin-Palomo}, \citenamefont {Weimann}, \citenamefont
  {Randel}, \citenamefont {Freude}, \citenamefont {Kippenberg},\ and\
  \citenamefont {Koos}}]{Trocha_2018}%
  \BibitemOpen
  \bibfield  {author} {\bibinfo {author} {\bibfnamefont {P.}~\bibnamefont
  {Trocha}}, \bibinfo {author} {\bibfnamefont {M.}~\bibnamefont {Karpov}},
  \bibinfo {author} {\bibfnamefont {D.}~\bibnamefont {Ganin}}, \bibinfo
  {author} {\bibfnamefont {M.~H.~P.}\ \bibnamefont {Pfeiffer}}, \bibinfo
  {author} {\bibfnamefont {A.}~\bibnamefont {Kordts}}, \bibinfo {author}
  {\bibfnamefont {S.}~\bibnamefont {Wolf}}, \bibinfo {author} {\bibfnamefont
  {J.}~\bibnamefont {Krockenberger}}, \bibinfo {author} {\bibfnamefont
  {P.}~\bibnamefont {Marin-Palomo}}, \bibinfo {author} {\bibfnamefont
  {C.}~\bibnamefont {Weimann}}, \bibinfo {author} {\bibfnamefont
  {S.}~\bibnamefont {Randel}}, \bibinfo {author} {\bibfnamefont
  {W.}~\bibnamefont {Freude}}, \bibinfo {author} {\bibfnamefont {T.~J.}\
  \bibnamefont {Kippenberg}}, \ and\ \bibinfo {author} {\bibfnamefont
  {C.}~\bibnamefont {Koos}},\ }\href {\doibase 10.1126/science.aao3924}
  {\bibfield  {journal} {\bibinfo  {journal} {Science}\ }\textbf {\bibinfo
  {volume} {359}},\ \bibinfo {pages} {887} (\bibinfo {year}
  {2018})}\BibitemShut {NoStop}%
\bibitem [{\citenamefont {Allen}\ \emph {et~al.}(1992)\citenamefont {Allen},
  \citenamefont {Beijersbergen}, \citenamefont {Spreeuw},\ and\ \citenamefont
  {Woerdman}}]{Allen_1992}%
  \BibitemOpen
  \bibfield  {author} {\bibinfo {author} {\bibfnamefont {L.}~\bibnamefont
  {Allen}}, \bibinfo {author} {\bibfnamefont {M.~W.}\ \bibnamefont
  {Beijersbergen}}, \bibinfo {author} {\bibfnamefont {R.~J.~C.}\ \bibnamefont
  {Spreeuw}}, \ and\ \bibinfo {author} {\bibfnamefont {J.~P.}\ \bibnamefont
  {Woerdman}},\ }\href {\doibase 10.1103/PhysRevA.45.8185} {\bibfield
  {journal} {\bibinfo  {journal} {Physical Review A}\ }\textbf {\bibinfo
  {volume} {45}},\ \bibinfo {pages} {8185} (\bibinfo {year}
  {1992})}\BibitemShut {NoStop}%
\bibitem [{\citenamefont {Barreiro}\ \emph {et~al.}(2006)\citenamefont
  {Barreiro}, \citenamefont {Tabosa}, \citenamefont {Failache},\ and\
  \citenamefont {Lezama}}]{Barreiro_2006}%
  \BibitemOpen
  \bibfield  {author} {\bibinfo {author} {\bibfnamefont {S.}~\bibnamefont
  {Barreiro}}, \bibinfo {author} {\bibfnamefont {J.~W.~R.}\ \bibnamefont
  {Tabosa}}, \bibinfo {author} {\bibfnamefont {H.}~\bibnamefont {Failache}}, \
  and\ \bibinfo {author} {\bibfnamefont {A.}~\bibnamefont {Lezama}},\ }\href
  {https://doi.org/10.1103%2Fphysrevlett.97.113601} {\bibfield  {journal}
  {\bibinfo  {journal} {Physical Review Letters}\ }\textbf {\bibinfo {volume}
  {97}} (\bibinfo {year} {2006})}\BibitemShut {NoStop}%
\bibitem [{\citenamefont {Lavery}\ \emph {et~al.}(2013)\citenamefont {Lavery},
  \citenamefont {Speirits}, \citenamefont {Barnett},\ and\ \citenamefont
  {Padgett}}]{Lavery_2013}%
  \BibitemOpen
  \bibfield  {author} {\bibinfo {author} {\bibfnamefont {M.~P.~J.}\
  \bibnamefont {Lavery}}, \bibinfo {author} {\bibfnamefont {F.~C.}\
  \bibnamefont {Speirits}}, \bibinfo {author} {\bibfnamefont {S.~M.}\
  \bibnamefont {Barnett}}, \ and\ \bibinfo {author} {\bibfnamefont {M.~J.}\
  \bibnamefont {Padgett}},\ }\href {\doibase 10.1126/science.1239936}
  {\bibfield  {journal} {\bibinfo  {journal} {Science}\ }\textbf {\bibinfo
  {volume} {341}},\ \bibinfo {pages} {537} (\bibinfo {year}
  {2013})}\BibitemShut {NoStop}%
\bibitem [{\citenamefont {Lavery}\ \emph {et~al.}(2014)\citenamefont {Lavery},
  \citenamefont {Barnett}, \citenamefont {Speirits},\ and\ \citenamefont
  {Padgett}}]{Lavery_2014}%
  \BibitemOpen
  \bibfield  {author} {\bibinfo {author} {\bibfnamefont {M.~P.~J.}\
  \bibnamefont {Lavery}}, \bibinfo {author} {\bibfnamefont {S.~M.}\
  \bibnamefont {Barnett}}, \bibinfo {author} {\bibfnamefont {F.~C.}\
  \bibnamefont {Speirits}}, \ and\ \bibinfo {author} {\bibfnamefont {M.~J.}\
  \bibnamefont {Padgett}},\ }\href {\doibase 10.1364/optica.1.000001}
  {\bibfield  {journal} {\bibinfo  {journal} {Optica}\ }\textbf {\bibinfo
  {volume} {1}},\ \bibinfo {pages} {1} (\bibinfo {year} {2014})}\BibitemShut
  {NoStop}%
\bibitem [{\citenamefont {Magana-Loaiza}\ \emph {et~al.}(2014)\citenamefont
  {Magana-Loaiza}, \citenamefont {Mirhosseini}, \citenamefont {Rodenburg},\
  and\ \citenamefont {Boyd}}]{Magana_2014}%
  \BibitemOpen
  \bibfield  {author} {\bibinfo {author} {\bibfnamefont {O.~S.}\ \bibnamefont
  {Magana-Loaiza}}, \bibinfo {author} {\bibfnamefont {M.}~\bibnamefont
  {Mirhosseini}}, \bibinfo {author} {\bibfnamefont {B.}~\bibnamefont
  {Rodenburg}}, \ and\ \bibinfo {author} {\bibfnamefont {R.~W.}\ \bibnamefont
  {Boyd}},\ }\href {https://link.aps.org/doi/10.1103/PhysRevLett.112.200401}
  {\bibfield  {journal} {\bibinfo  {journal} {Physical Review Letters}\
  }\textbf {\bibinfo {volume} {112}} (\bibinfo {year} {2014})}\BibitemShut
  {NoStop}%
\bibitem [{\citenamefont {D'Ambrosio}\ \emph {et~al.}(2013)\citenamefont
  {D'Ambrosio}, \citenamefont {Spagnolo}, \citenamefont {Re}, \citenamefont
  {Slussarenko}, \citenamefont {Li}, \citenamefont {Kwek}, \citenamefont
  {Marrucci}, \citenamefont {Walborn}, \citenamefont {Aolita},\ and\
  \citenamefont {Sciarrino}}]{D_Ambrosio_2013}%
  \BibitemOpen
  \bibfield  {author} {\bibinfo {author} {\bibfnamefont {V.}~\bibnamefont
  {D'Ambrosio}}, \bibinfo {author} {\bibfnamefont {N.}~\bibnamefont
  {Spagnolo}}, \bibinfo {author} {\bibfnamefont {L.~D.}\ \bibnamefont {Re}},
  \bibinfo {author} {\bibfnamefont {S.}~\bibnamefont {Slussarenko}}, \bibinfo
  {author} {\bibfnamefont {Y.}~\bibnamefont {Li}}, \bibinfo {author}
  {\bibfnamefont {L.~C.}\ \bibnamefont {Kwek}}, \bibinfo {author}
  {\bibfnamefont {L.}~\bibnamefont {Marrucci}}, \bibinfo {author}
  {\bibfnamefont {S.~P.}\ \bibnamefont {Walborn}}, \bibinfo {author}
  {\bibfnamefont {L.}~\bibnamefont {Aolita}}, \ and\ \bibinfo {author}
  {\bibfnamefont {F.}~\bibnamefont {Sciarrino}},\ }\href
  {https://doi.org/10.1038%2Fncomms3432} {\bibfield  {journal} {\bibinfo
  {journal} {Nature Communications}\ }\textbf {\bibinfo {volume} {4}} (\bibinfo
  {year} {2013})}\BibitemShut {NoStop}%
\bibitem [{\citenamefont {Torner}\ \emph {et~al.}(2005)\citenamefont {Torner},
  \citenamefont {Torres},\ and\ \citenamefont {Carrasco}}]{Torner_2005}%
  \BibitemOpen
  \bibfield  {author} {\bibinfo {author} {\bibfnamefont {L.}~\bibnamefont
  {Torner}}, \bibinfo {author} {\bibfnamefont {J.~P.}\ \bibnamefont {Torres}},
  \ and\ \bibinfo {author} {\bibfnamefont {S.}~\bibnamefont {Carrasco}},\
  }\href {\doibase 10.1364/opex.13.000873} {\bibfield  {journal} {\bibinfo
  {journal} {Optics Express}\ }\textbf {\bibinfo {volume} {13}},\ \bibinfo
  {pages} {873} (\bibinfo {year} {2005})}\BibitemShut {NoStop}%
\bibitem [{\citenamefont {Magana-Loaiza}\ and\ \citenamefont
  {Boyd}(2019)}]{Omar_2019}%
  \BibitemOpen
  \bibfield  {author} {\bibinfo {author} {\bibfnamefont {O.~S.}\ \bibnamefont
  {Magana-Loaiza}}\ and\ \bibinfo {author} {\bibfnamefont {R.~W.}\ \bibnamefont
  {Boyd}},\ }\href {http://iopscience.iop.org/10.1088/1361-6633/ab5005}
  {\bibfield  {journal} {\bibinfo  {journal} {Reports on Progress in Physics}\
  } (\bibinfo {year} {2019})}\BibitemShut {NoStop}%
\bibitem [{\citenamefont {Yang}\ \emph {et~al.}(2017)\citenamefont {Yang},
  \citenamefont {Magana-Loaiza}, \citenamefont {Mirhosseini}, \citenamefont
  {Zhou}, \citenamefont {Gao}, \citenamefont {Gao}, \citenamefont {Rafsanjani},
  \citenamefont {Long},\ and\ \citenamefont {Boyd}}]{Yang_2017}%
  \BibitemOpen
  \bibfield  {author} {\bibinfo {author} {\bibfnamefont {Z.}~\bibnamefont
  {Yang}}, \bibinfo {author} {\bibfnamefont {O.~S.}\ \bibnamefont
  {Magana-Loaiza}}, \bibinfo {author} {\bibfnamefont {M.}~\bibnamefont
  {Mirhosseini}}, \bibinfo {author} {\bibfnamefont {Y.~Y.}\ \bibnamefont
  {Zhou}}, \bibinfo {author} {\bibfnamefont {B.~S.}\ \bibnamefont {Gao}},
  \bibinfo {author} {\bibfnamefont {L.}~\bibnamefont {Gao}}, \bibinfo {author}
  {\bibfnamefont {S.~M.~H.}\ \bibnamefont {Rafsanjani}}, \bibinfo {author}
  {\bibfnamefont {G.~L.}\ \bibnamefont {Long}}, \ and\ \bibinfo {author}
  {\bibfnamefont {R.~W.}\ \bibnamefont {Boyd}},\ }\href
  {https://doi.org/10.1038%2Flsa.2017.13} {\bibfield  {journal} {\bibinfo
  {journal} {Light: Science {\&} Applications}\ }\textbf {\bibinfo {volume}
  {6}} (\bibinfo {year} {2017})}\BibitemShut {NoStop}%
\bibitem [{\citenamefont {Uribe-Patarroyo}\ \emph {et~al.}(2013)\citenamefont
  {Uribe-Patarroyo}, \citenamefont {Fraine}, \citenamefont {Simon},
  \citenamefont {Minaeva},\ and\ \citenamefont
  {Sergienko}}]{Uribe_Patarroyo_2013}%
  \BibitemOpen
  \bibfield  {author} {\bibinfo {author} {\bibfnamefont {N.}~\bibnamefont
  {Uribe-Patarroyo}}, \bibinfo {author} {\bibfnamefont {A.}~\bibnamefont
  {Fraine}}, \bibinfo {author} {\bibfnamefont {D.~S.}\ \bibnamefont {Simon}},
  \bibinfo {author} {\bibfnamefont {O.}~\bibnamefont {Minaeva}}, \ and\
  \bibinfo {author} {\bibfnamefont {A.~V.}\ \bibnamefont {Sergienko}},\ }\href
  {https://doi.org/10.1103%2Fphysrevlett.110.043601} {\bibfield  {journal}
  {\bibinfo  {journal} {Physical Review Letters}\ }\textbf {\bibinfo {volume}
  {110}} (\bibinfo {year} {2013})}\BibitemShut {NoStop}%
\bibitem [{\citenamefont {Zhang}\ \emph {et~al.}(2018)\citenamefont {Zhang},
  \citenamefont {Gao}, \citenamefont {Zhang}, \citenamefont {He}, \citenamefont
  {Xu}, \citenamefont {Fickler},\ and\ \citenamefont {Chen}}]{Zhang_2018}%
  \BibitemOpen
  \bibfield  {author} {\bibinfo {author} {\bibfnamefont {W.}~\bibnamefont
  {Zhang}}, \bibinfo {author} {\bibfnamefont {J.}~\bibnamefont {Gao}}, \bibinfo
  {author} {\bibfnamefont {D.}~\bibnamefont {Zhang}}, \bibinfo {author}
  {\bibfnamefont {Y.}~\bibnamefont {He}}, \bibinfo {author} {\bibfnamefont
  {T.}~\bibnamefont {Xu}}, \bibinfo {author} {\bibfnamefont {R.}~\bibnamefont
  {Fickler}}, \ and\ \bibinfo {author} {\bibfnamefont {L.}~\bibnamefont
  {Chen}},\ }\href {https://doi.org/10.1103%2Fphysrevapplied.10.044014}
  {\bibfield  {journal} {\bibinfo  {journal} {Physical Review Applied}\
  }\textbf {\bibinfo {volume} {10}} (\bibinfo {year} {2018})}\BibitemShut
  {NoStop}%
\bibitem [{\citenamefont {Qiu}\ \emph {et~al.}(2019)\citenamefont {Qiu},
  \citenamefont {Zhang}, \citenamefont {Zhang},\ and\ \citenamefont
  {Chen}}]{Qiu_2019}%
  \BibitemOpen
  \bibfield  {author} {\bibinfo {author} {\bibfnamefont {X.}~\bibnamefont
  {Qiu}}, \bibinfo {author} {\bibfnamefont {D.}~\bibnamefont {Zhang}}, \bibinfo
  {author} {\bibfnamefont {W.}~\bibnamefont {Zhang}}, \ and\ \bibinfo {author}
  {\bibfnamefont {L.}~\bibnamefont {Chen}},\ }\href
  {https://doi.org/10.1103%2Fphysrevlett.122.123901} {\bibfield  {journal}
  {\bibinfo  {journal} {Physical Review Letters}\ }\textbf {\bibinfo {volume}
  {122}} (\bibinfo {year} {2019})}\BibitemShut {NoStop}%
\bibitem [{\citenamefont {Kulkarni}\ \emph {et~al.}(2017)\citenamefont
  {Kulkarni}, \citenamefont {Sahu}, \citenamefont {Maga{\~{n}}a-Loaiza},
  \citenamefont {Boyd},\ and\ \citenamefont {Jha}}]{Kulkarni_2017}%
  \BibitemOpen
  \bibfield  {author} {\bibinfo {author} {\bibfnamefont {G.}~\bibnamefont
  {Kulkarni}}, \bibinfo {author} {\bibfnamefont {R.}~\bibnamefont {Sahu}},
  \bibinfo {author} {\bibfnamefont {O.~S.}\ \bibnamefont
  {Maga{\~{n}}a-Loaiza}}, \bibinfo {author} {\bibfnamefont {R.~W.}\
  \bibnamefont {Boyd}}, \ and\ \bibinfo {author} {\bibfnamefont {A.~K.}\
  \bibnamefont {Jha}},\ }\href {https://doi.org/10.1038%2Fs41467-017-01215-x}
  {\bibfield  {journal} {\bibinfo  {journal} {Nature Communications}\ }\textbf
  {\bibinfo {volume} {8}} (\bibinfo {year} {2017})}\BibitemShut {NoStop}%
\bibitem [{\citenamefont {Schulze}\ \emph {et~al.}(2015)\citenamefont
  {Schulze}, \citenamefont {Roux}, \citenamefont {Dudley}, \citenamefont {Rop},
  \citenamefont {Duparr{\'{e}}},\ and\ \citenamefont {Forbes}}]{Schulze_2015}%
  \BibitemOpen
  \bibfield  {author} {\bibinfo {author} {\bibfnamefont {C.}~\bibnamefont
  {Schulze}}, \bibinfo {author} {\bibfnamefont {F.~S.}\ \bibnamefont {Roux}},
  \bibinfo {author} {\bibfnamefont {A.}~\bibnamefont {Dudley}}, \bibinfo
  {author} {\bibfnamefont {R.}~\bibnamefont {Rop}}, \bibinfo {author}
  {\bibfnamefont {M.}~\bibnamefont {Duparr{\'{e}}}}, \ and\ \bibinfo {author}
  {\bibfnamefont {A.}~\bibnamefont {Forbes}},\ }\href
  {https://doi.org/10.1103%2Fphysreva.91.043821} {\bibfield  {journal}
  {\bibinfo  {journal} {Physical Review A}\ }\textbf {\bibinfo {volume} {91}}
  (\bibinfo {year} {2015})}\BibitemShut {NoStop}%
\bibitem [{\citenamefont {Webster}\ \emph {et~al.}(2017)\citenamefont
  {Webster}, \citenamefont {Rosales-Guzm{\'{a}}n},\ and\ \citenamefont
  {Forbes}}]{Webster_2017}%
  \BibitemOpen
  \bibfield  {author} {\bibinfo {author} {\bibfnamefont {J.}~\bibnamefont
  {Webster}}, \bibinfo {author} {\bibfnamefont {C.}~\bibnamefont
  {Rosales-Guzm{\'{a}}n}}, \ and\ \bibinfo {author} {\bibfnamefont
  {A.}~\bibnamefont {Forbes}},\ }\href {\doibase 10.1364/ol.42.000675}
  {\bibfield  {journal} {\bibinfo  {journal} {Optics Letters}\ }\textbf
  {\bibinfo {volume} {42}},\ \bibinfo {pages} {675} (\bibinfo {year}
  {2017})}\BibitemShut {NoStop}%
\bibitem [{\citenamefont {Hsu}\ \emph {et~al.}(1982)\citenamefont {Hsu},
  \citenamefont {Arsenault},\ and\ \citenamefont {April}}]{Hsu_1982_rotation}%
  \BibitemOpen
  \bibfield  {author} {\bibinfo {author} {\bibfnamefont {Y.-N.}\ \bibnamefont
  {Hsu}}, \bibinfo {author} {\bibfnamefont {H.~H.}\ \bibnamefont {Arsenault}},
  \ and\ \bibinfo {author} {\bibfnamefont {G.}~\bibnamefont {April}},\ }\href
  {\doibase 10.1364/ao.21.004012} {\bibfield  {journal} {\bibinfo  {journal}
  {Applied Optics}\ }\textbf {\bibinfo {volume} {21}},\ \bibinfo {pages} {4012}
  (\bibinfo {year} {1982})}\BibitemShut {NoStop}%
\bibitem [{Note1()}]{Note1}%
  \BibitemOpen
  \bibinfo {note} {As for light spot with radius of micron scale and commercial
  CCD or CMOS whose pixel size is also several micrometers, one may need a
  microscope system with over 300$\times $ magnification to make the errors
  result from finite resolution smaller than $10^{-10}$}\BibitemShut {NoStop}%
\bibitem [{\citenamefont {Liang}\ \emph {et~al.}(2018)\citenamefont {Liang},
  \citenamefont {Zhu},\ and\ \citenamefont {Wang}}]{Liang_2018}%
  \BibitemOpen
  \bibfield  {author} {\bibinfo {author} {\bibfnamefont {J.}~\bibnamefont
  {Liang}}, \bibinfo {author} {\bibfnamefont {L.}~\bibnamefont {Zhu}}, \ and\
  \bibinfo {author} {\bibfnamefont {L.~V.}\ \bibnamefont {Wang}},\ }\href
  {https://doi.org/10.1038%2Fs41377-018-0044-7} {\bibfield  {journal} {\bibinfo
   {journal} {Light: Science {\&} Applications}\ }\textbf {\bibinfo {volume}
  {7}} (\bibinfo {year} {2018})}\BibitemShut {NoStop}%
\bibitem [{\citenamefont {Hsu}\ and\ \citenamefont
  {Arsenault}(1982)}]{Hsu_1982}%
  \BibitemOpen
  \bibfield  {author} {\bibinfo {author} {\bibfnamefont {Y.-N.}\ \bibnamefont
  {Hsu}}\ and\ \bibinfo {author} {\bibfnamefont {H.~H.}\ \bibnamefont
  {Arsenault}},\ }\href {\doibase 10.1364/ao.21.004016} {\bibfield  {journal}
  {\bibinfo  {journal} {Applied Optics}\ }\textbf {\bibinfo {volume} {21}},\
  \bibinfo {pages} {4016} (\bibinfo {year} {1982})}\BibitemShut {NoStop}%
\bibitem [{\citenamefont {Harmany}\ \emph {et~al.}(2012)\citenamefont
  {Harmany}, \citenamefont {Marcia},\ and\ \citenamefont
  {Willett}}]{Harmany_2012}%
  \BibitemOpen
  \bibfield  {author} {\bibinfo {author} {\bibfnamefont {Z.~T.}\ \bibnamefont
  {Harmany}}, \bibinfo {author} {\bibfnamefont {R.~F.}\ \bibnamefont {Marcia}},
  \ and\ \bibinfo {author} {\bibfnamefont {R.~M.}\ \bibnamefont {Willett}},\
  }\href {https://doi.org/10.1109%2Ftip.2011.2168410} {\bibfield  {journal}
  {\bibinfo  {journal} {{IEEE} Transactions on Image Processing}\ }\textbf
  {\bibinfo {volume} {21}},\ \bibinfo {pages} {1084} (\bibinfo {year}
  {2012})}\BibitemShut {NoStop}%
\bibitem [{\citenamefont {Bartoszyǹski}\ and\ \citenamefont
  {Niewiadomska-Bugaj}(2007)}]{Bartoszy_ski_2007}%
  \BibitemOpen
  \bibfield  {author} {\bibinfo {author} {\bibfnamefont {R.}~\bibnamefont
  {Bartoszyǹski}}\ and\ \bibinfo {author} {\bibfnamefont {M.}~\bibnamefont
  {Niewiadomska-Bugaj}},\ }\href {\doibase 10.1002/9780470191590} {\emph
  {\bibinfo {title} {Probability and Statistical Inference, Second Edition}}}\
  (\bibinfo  {publisher} {John Wiley {\&} Sons, Inc.},\ \bibinfo {year}
  {2007})\BibitemShut {NoStop}%
\bibitem [{\citenamefont {Devore}\ and\ \citenamefont
  {Berk}(2012)}]{Devore_2012}%
  \BibitemOpen
  \bibfield  {author} {\bibinfo {author} {\bibfnamefont {J.~L.}\ \bibnamefont
  {Devore}}\ and\ \bibinfo {author} {\bibfnamefont {K.~N.}\ \bibnamefont
  {Berk}},\ }\href {\doibase 10.1007/978-1-4614-0391-3} {\emph {\bibinfo
  {title} {Modern Mathematical Statistics with Applications}}}\ (\bibinfo
  {publisher} {Springer New York},\ \bibinfo {year} {2012})\BibitemShut
  {NoStop}%
\end{thebibliography}%
\end{document}